\documentclass[letterpaper]{aa}
\usepackage{txfonts}
\usepackage{natbib}
\usepackage{graphicx}
\bibpunct{(}{)}{;}{a}{}{,}

\begin{document}

\title{The influence of non-isotropic scattering of thermal radiation on spectra of brown dwarfs and hot exoplanets}
\author{R.J.~de Kok\inst{\ref{int1}} \and Ch.~Helling\inst{\ref{int2}} \and D.M.~Stam\inst{\ref{int1}} \and P.~Woitke\inst{\ref{int3}\ref{int2}\ref{int4}} \and S.~Witte\inst{\ref{int5}}}
\institute{SRON Netherlands Institute for Space Research, Sorbonnelaan 2, 3584 CA Utrecht, the Netherlands \email{R.J.de.Kok@sron.nl} \label{int1} \and SUPA, School of Physics \& Astronomy, University of St.~Andrews, North Haugh, St.~Andrews KY16 9SS, UK \label{int2} \and University of Vienna, Dpet.~of Astronomy, T{\"u}rkenschanzstr.~17, A-1180 Vienna, Austria \label{int3} \and UK Astronomy Technology Centre, Royal Observatory, Edinburgh, Blackford Hill, Edinburgh EH9 3HJ, UK \label{int4} \and Hamburger Sternwarte, Gojenbergsweg 112, 21029 Hamburg, Germany, \label{int5}}
\date{}

\abstract
{Currently, the thermal emission from exoplanets can be measured with either direct imaging or secondary eclipse measurements of transiting exoplanets. Most of these measurements are taken at near-infrared wavelengths, where the thermal emission of these planets peaks. Cool brown dwarfs, covering a similar temperature range, are also mostly characterised using near-infrared spectra.}{We aim to show how thermal radiation in brown dwarf and exoplanet atmospheres can be scattered by clouds and haze and to investigate how the thermal emission spectrum is changed when different assumptions in the radiative transfer modelling are made.}{We calculate near-infrared thermal emission spectra using a doubling-adding radiative transfer code, which includes scattering by clouds and haze. Initial temperature profiles and cloud optical depths are taken from the \textsc{drift-phoenix} brown dwarf model.}{As is well known, cloud particles change the spectrum compared to the same atmosphere with the clouds ignored. The clouds reduce fluxes in the near-infrared spectrum and make it redder than for the clear sky case. We also confirm that not including scattering in the spectral calculations can result in errors on the spectra of many tens of percent, both in magnitude and in variations with wavelength. This is especially apparent for particles that are larger than the wavelength and only have little iron in them. Scattering particles will show deeper absorption features than absorbing (e.g.~iron) particles and scattering and particle size will also affect the calculated infrared colours. Large particles also tend to be strongly forward-scattering, and we show that assuming isotropic scattering in this case also leads to very large errors in the spectrum. Thus, care must be taken in the choice of radiative transfer method for heat balance or spectral calculations when clouds are present in the atmosphere. Besides the choice of radiative transfer method, the type of particles that are predicted by models will change conclusions about e.g.~infrared colours and trace gas abundances. As a result, knowledge of the scattering properties of the clouds is essential when deriving temperature profiles or gas abundances from direct infrared observations of exoplanets or brown dwarfs and from secondary eclipse measurements of transiting exoplanets, since scattering clouds will change the depth of gas absorption features, among other things. Thus, ignoring the presence of clouds can yield retrieved properties that differ significantly from the real atmospheric properties.}{}


\titlerunning{Scattering of thermal radiation on brown dwarfs and hot exoplanets}
\authorrunning{de Kok et al.}

\maketitle

\section{Introduction}

In recent years characterisating the atmospheres of exoplanets has started to become possible using spectroscopy. The planets for which measurements at multiple wavelengths are available now are either large transiting planets \citep[e.g.][and references therein]{dem09}, or large planets far from their star \citep[e.g.][]{pat10,bow10,jan10}. For these planets the thermal emission can be measured, either from its secondary eclipse in the former case, or by direct imaging in the latter case. These planets are either very close to their star (transit measurements) or very young (direct detection), meaning that they tend to be hot and their thermal emission peaks at near-infrared wavelengths. Therefore, most measurements of the thermal emission, from direct imaging or secondary eclipse, are taken at near-infrared wavelengths. Along with reflecting starlight, infrared light from the planet itself can be scattered when clouds or hazes are present in the planet's atmosphere. This scattering of the light can significantly alter the heat balance in the atmosphere and change the thermal emission spectrum of the planet compared to the case without scattering. This scattering behaviour is evident for thermal emission from, e.g.,~Venus \citep{gri93,tsa08} and Jupiter \citep{car93} and can be expected to be pronounced at the short infrared wavelengths where hot exoplanets emit most of their thermal radiation.

Clouds have already been detected on brown dwarfs  \citep{tsu96,leg98,all01,ack01,cus06,burg09} and have been suggested for young planets orbiting at large orbital distances \citep[e.g.][]{bow10,jan10,bon10}. Also, transit tranmission measurements of hot exoplanets have revealed hazes \citep{lec08,pon08,sin09,woo11}. However, comparisons of atmospheric models with infrared secondary eclipse measurements that probe the dayside thermal emission of transiting exoplanets \citep[e.g.][]{bur08,knu09,mad09,swai09,cro10} have so far neglected the possible presence of clouds or haze. 

At the relatively cool temperatures of substellar atmospheres, dust and cloud formation is predicted to significantly affect the thermal structure of the atmosphere. Some numerical algorithms are available that predict cloud properties in brown dwarfs, given a certain set of input parameters, such as the effective temperature and the gravity. These models probe temperatures similar to hot giant exoplanets. The available models can be divided into the ones that use a kinetic approach for dust formation \citep[e.g.][]{woi03,woi04,hel08b} and the ones that assume phase equilibrium \citep[e.g.][]{tsu96,ack01,all01,bar01,bar05,bur06}. A recent comparison shows that the resulting atmospheres can differ substantially between the different brown dwarf cloud models, given the same set of input parameters \citep{hel08}, although the models that include both chemistry and opacity in the cloud formation show a relatively good agreement (F.~Allard, pers.~comm.). 

The radiative transfer methods that are used to calculate spectra (SEDs) and heat balance in the available brown dwarf models also differ. The equation of radiative transfer in a plane-parallel atmosphere can be written as \citep[e.g.][]{cha60,hub03}:

\begin{equation}
\mu \frac{dI_\nu}{d \tau_\nu} = I_\nu - S_\nu
\label{eq.rad}
\end{equation}

Here, $\mu$ is the directional cosine, $I_\nu$ is the wavelength-dependent radiance, $\tau_\nu$ is the optical depth, and $S_\nu$ is the source function. In a planetary atmosphere all these normally vary with altitude or pressure and Eq.~\ref{eq.rad} has to be solved taking into account the entire atmosphere to yield the radiation that is emitted into space. The source function generally has contributions from both thermal emission of the local atmosphere itself and from scattering of radiation coming from other parts of the atmosphere. In local thermal equilibrium, the emission source function is the Planck function for black-body radiation, $B_\nu$. The scattering source function can be written as:

\begin{equation}
S_{\nu,s} = \frac{1}{4 \pi} \int P_\nu I_\nu d\Omega
\label{eq.source}
\end{equation}

In the scattering source function the radiation $I_\nu$ coming in from different directions are integrated over solid angle $\Omega$ and are weighted according to the phase function $P_\nu$. The two contributions to the total source function are then weighted according to the single-scattering albedo $\tilde{\omega}_0$:

\begin{equation}
S_\nu = (1-\tilde{\omega}_0)B_\nu + \tilde{\omega}_0 S_{\nu,s}
\label{eq.source2}
\end{equation}

The single-scattering albedo is the fraction of the light that is scattered when initially reaching a particle or layer in the atmosphere. The remaining fraction is absorbed. The term single-scattering albedo should not be confused with `single-scattering' methods of radiative transfer, where only one order of scattering is taken into account when solving Eq.~1. In the method used here, and also in the literature referenced here, multiple scattering radiative transfer is used, i.e.~light is scattered more than once between different particles or atmospheric layers. In stellar literature the collisional destruction probability, $\epsilon = 1 - \tilde{\omega}_0$, is commonly used instead of the single-scattering albedo.

The equation of radiative transfer is solved differently in different brown dwarf models. The \textsc{phoenix} model \citep{hau92,all95,hau06} increases efficiency of the calculations by assuming isotropic scattering, i.e.~the phase function in Eq.~\ref{eq.source} distributes the light equally in all directions (P.~Hauschildt, pers.~comm.). The equation of radiative transfer is then solved iteratively at multiple angles through the atmosphere, using a method based on approximate $\Lambda$-iteration \citep[e.g.][]{hub03}.  \citet{bur06,bur08} use a version of the \textsc{tlusty} code \citep{hub95} to iteratively compute the radiative transfer within the atmosphere. There radiation is calculated at multiple angles, with the phase function parameterised in terms of the asymmetry parameter $g$ (A. Burrows, pers.~comm.). A value of $g=1$ would indicate perfectly forward-scattering particles, whereas $g=0$ indicates isotropic scattering.\citet{ack01} use a two-step approach in calculating the spectrum and radiation field. Initially, they calculate the internal radiation field
for the source function ($I_\nu$ in Eq.~\ref{eq.source}) at eight  
different angles in each atmospheric layer.  For the scattering term only they  
use the two-stream method of \citet{too89} (M. Marley,
pers.~comm.) as the background source from which the scattering from  
particulates is computed. The two-stream approximation gives an exact  
solution for the case where radiation only consists of an up-going stream and a
down-going stream. In this case the phase function can again be
described using $g$, at the cost of a drop in accuracy. The two-stream  
source function method itself generally gives error less than $\sim$10-20\% for a single layer \citep{too89}. With the source function fully described, Eq.~1 can simply be integrated numerically to yield to radiation at the top of the
atmosphere at multiple angles. Because Ackerman and Marley only use the
two-stream method to compute the scattering source term
and not for calculating the final spectrum and radiation field, their method is
presumably somewhat more accurate than the two-stream method of
\citet{too89}.

Although scattering of the thermal emission is included in the previously discussed models, the various assumptions or approximations that are currently used to model the scattering may lead to significant errors in the calculated spectra and hence in errors on atmospheric properties derived from observations. Here, we will calculate near-infrared thermal emission spectra for specific model brown dwarf atmospheres, which are similar in temperature regime as hot exoplanets. Our calculations accurately include single and multiple scattering of the thermal radiation using scattering particles with angularly resolved phase functions.  Hence, we will make an assessment of the effect of scattering by clouds and haze on the thermal emission spectra of hot exoplanet atmospheres  and can compare spectra with different scattering assumptions in isolation, instead of comparing spectra from different cloud models that couple physics, chemistry and radiative transfer.

%%%%%%%%%%%%%%%%%%%%%%%%%%%%%%%%%%%%%%%%%%%%%%%%%%%%%%%%%%%%%%%%%%%%%%%%%%%%%%%%%%%%%%%%%%%%%%%%%

\section{Model atmospheres}

In this paper, we do not self-consistently calculate temperature and composition profiles of 
the model atmospheres with our radiative transfer code, but instead use model atmospheres 
calculated with the self-consistent \textsc{drift-phoenix} code. 
With the \textsc{drift-phoenix} code, atmospheric cloud particles are formed when rising trace 
gases cool down and seeds are formed. The particles are transported 
through the atmosphere, accumulating more material and/or evaporating again. 
The resulting cloud particles usually consist of mixtures of several solids and are therefore 
called `dirty' particles. Details about the actual computations performed by \textsc{drift-phoenix} can be found in \citet{hel08a,hel08b,wit09}.

\begin{figure}[htp]
\centering
\resizebox{\hsize}{!}{\includegraphics{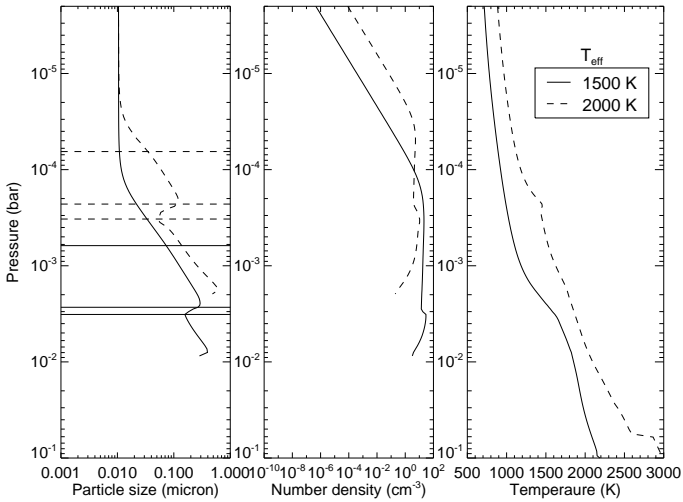}}
\caption{Mean particle sizes,
         number densities and temperature profiles for our two 
         model atmospheres. In the atmosphere with $T_\mathrm{eff}=1500$~K ($T_\mathrm{eff}=2000$~K), 
         particle condensation starts at a pressure of 10$^{-2}$~bar (2$\cdot$~10$^{-3}$~bar). 
         The horizontal lines in the left panel indicate the boundaries of 
         the four particle types that we define for our radiative transfer calculations 
         (see text and Fig.~\ref{fig.volp}), which are different for the two atmospheres.
         }
\label{fig.atm}
\end{figure}

In this paper, we use two model atmospheres that are calculated using \textsc{drift-phoenix}, 
assuming solar metallicity 
and a gravitational acceleration of 10 ms$^{-2}$. The model atmospheres only differ in the given
effective temperature (T$_{\mathrm{eff}}$=1500~K and 2000~K, respectively).  
The calculated temperatures profiles and cloud particle sizes and number densities for the 
two model atmospheres are shown in Fig.~\ref{fig.atm}. 
We use the vertical profiles of the temperature, trace gas mixing ratios (defined as the trace gas number densities divided by the total gas number density), cloud
particle number densities and microphysical properties (particle size and composition) versus pressure as input for 
our radiative transfer calculations. The size distributions of the particles in \textsc{drift-phoenix} are described by moments \citep{hel08b}, but here we will only use the mean particle size as a function of pressure from \textsc{drift-phoenix}.

The vertical profiles provided by \textsc{drift-phoenix} consist of $\sim$400 homogeneous layers.
To reduce radiative transfer computation times, we reduce the number of homogeneous layers to 40
by interpolating the temperatures, pressures and gas abundances, and binning the optical thicknesses of the clouds. This significantly reduces the number of layers, while still resolving the different cloud layers in the atmosphere.
In addition, \textsc{drift-phoenix} predicts particles that have compositions and sizes
that that are altitude dependent. As a result, the absorption and scattering properties 
(single scattering albedo, extinction cross--section and scattering matrix)
of the particles depend not only on the wavelength but also on the altitude.
Taking these variations fully into account would require wavelength dependent
calculations of the optical properties of 80 different particles (40 layers and
two model atmospheres). Instead, we define four different particles for each model
atmosphere (thus, eight in total), corresponding to four altitude ranges in each model
atmosphere. Within each altitude range, the particle type is the same, only the
particles' number density varies between the layers.
Considering the almost total lack of constraints on clouds on exoplanets 
from current observations and condensation models, this reduction of details seems
reasonable for this exploratory study that focuses on the radiative transfer in a 
given model atmosphere. 

\begin{figure}[htp]
\centering
\resizebox{\hsize}{!}{\includegraphics{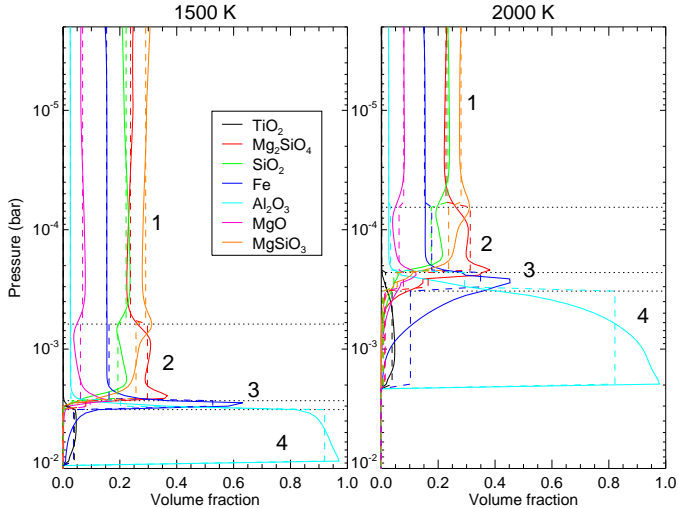}}
\caption{The volume fraction of the different constituents of the dirty particles produced
         by \textsc{drift-phoenix} for the two model atmospheres. The horizontal dotted
         lines indicate altitudes where the composition of the particles changes rapidly,
         which is where we have chosen the boundaries between the four different particle
         types. The particle types can be described as: 
         1 -- high silicate haze; 2 -- low silicate haze; 3 -- iron cloud; 4 -- Al$_2$O$_3$ cloud. 
         The altitude independent composition that we assume for each particle type 
         is indicated by vertical dashed, coloured lines (see Table~2).}
\label{fig.volp}
\end{figure}

Our different particle types are based on the changes of particle composition in the atmosphere. 
All particles produces by \textsc{drift-phoenix} are altitude dependent mixtures of various solids. Figure~\ref{fig.volp} shows how the composition of 
the particles varies across the two model atmospheres. As can be seen, across large regions of each 
atmosphere, the particle composition only varies slightly with pressure. At a few altitudes,
however, the particle composition 
changes rapidly. We choose these altitudes as the boundaries between our particle types. 
Based on these boundaries, the particle types in the two model atmospheres have the same
composition but different sizes (see below). We distinguish the following four particle
types (from the top of the atmosphere to the bottom), named after their main constituent: 
a high silicate haze,
a low silicate haze, an iron cloud, and an aluminium oxide (Al$_2$O$_3$) cloud.

Although particle composition stays roughly constant with altitude across the four regions in each
model atmosphere, particle size does not (see Fig.~\ref{fig.atm}). Since particle size is very 
important in determining the scattering properties of particles, assuming an altitude independent
particle size across a region might have some consequences for our radiative transfer results.
Fortunately, the particles that \textsc{drift-phoenix} produces tend to be very small, 
especially in the upper atmosphere (above $\sim$1 mbar), where particle size changes most rapidly. 
When particles are much smaller than the wavelength, however, their extinction cross-section 
scales with the cube of their radius, while its spectral dependence does not change \citep{han03}. 
Hence, for the lower and upper silicate haze, we take into account the altitude dependent 
particle size in the calculation of a layer's optical thickness, but leave the spectral 
dependence unchanged. 
Lower in the atmosphere, the particles are larger, but their size also changes less with altitude. 
Hence, for the atmospheric layers containing the iron and Al$_2$O$_3$ clouds, we assume a
constant (the mean) particle radius. The assumed particle sizes used in the Mie calculations are given in Table~1. These coincide with the average of the mean particle size over the pressure range considered. Table~2 summarises the composition of the dirty particles for each of the particle types.

\begin{table}
\begin{tabular}{lcc}
\hline
$T_\mathrm{eff}$: & 1500 K & 2000 K \\
\hline
High silicate haze & 0.05 & 0.03 \\
Low silicate haze & 0.07 & 0.10 \\
Iron cloud & 0.22 & 0.07 \\
Al$_2$O$_3$ cloud & 0.26 & 0.25 \\
\hline
\end{tabular}
\caption{Assumed particle sizes ($\mu$m) for the four different particle types 
         in the two model atmospheres.}
\end{table}

\begin{table}
\begin{tabular}{lcccc}
\hline
 & High Si haze & Low Si Haze & Iron cloud & Al$_2$O$_3$ cloud \\
 \hline
TiO$_2$ &	0 \emph{(0)}&     0 \emph{(0)}&   1 \emph{(2)}&	   4 \emph{(4)}\\
Mg$_2$SiO$_4$ & 24 \emph{(23)}&   30 \emph{(31)}& 8 \emph{(16)}&   0 \emph{(1)}\\
SiO$_2$ &	22 \emph{(23)}&   19 \emph{(18)}& 3 \emph{(4)}&   0 \emph{(1)}\\
Fe &		15 \emph{(15)}&   16 \emph{(18)}& 53 \emph{(35)}&   4 \emph{(10)}\\
Al$_2$O$_3$ &	3 \emph{(3)}&     3 \emph{(3)}&   20 \emph{(29)}&   92 \emph{(82)}\\
MgO &		7 \emph{(8)}&     6 \emph{(6)}&   10 \emph{(8)}&	   0 \emph{(2)}\\
 MgSiO$_3$ &	29 \emph{(28)}&   26 \emph{(24)}& 5 \emph{(6)}&   0 \emph{(0)}\\
\hline
\end{tabular}
\caption{Assumed volume fraction (\%) of constituents of the dirty particles for T$_\mathrm{eff}$ = 1500 \emph{(2000)} K.}
\end{table}

%%%%%%%%%%%%%%%%%%%%%%%%%%%%%%%%%%%%%%%%%%%%%%%%%%%%%%%%%%%%%%%%%%%%%%%%%%%%%%%%%%%%%%%%%%%%%%%%%

\section{Radiative transfer algorithm}

Using the model atmospheres produced by \textsc{drift-phoenix} as input, we calculate 
disc-integrated (spatially unresolved) near-infrared flux spectra with a doubling-adding algorithm.
The doubling-adding method is 
often used in calculations of sunlight that is reflected by and/or transmitted through 
planetary atmospheres, see e.g. \citet{han74} and \citet{deh87}. It allows the accurate 
solving of the equation of radiative transfer along paths in a model atmosphere that 
consists of a stack of plane-parallel, homogeneous layers containing scattering and/or 
absorbing gaseous molecules and/or aerosol particles. Vertically inhomogeneous model atmospheres 
are created by stacking different homogeneous layers. As described in Sect.~2.1, our model atmospheres consist of 40 layers.

As described in detail by e.g. \citet{deh87}, the doubling-adding method makes use of the
fact that if one knows the reflection and transmission properties of two adjacent atmospheric
layers, one can straightforwardly calculate these properties for
the combined layer. For each atmospheric layer, the calculation of its 
reflection and transmission properties starts with calculating them analytically
for a horizontal slice of the layer with a very small optical thickness. 
Then, using the so-called 'doubling'-equations \citep[see][]{deh87}, the properties
of a layer with twice the inital optical thickness are calculated. This is repeated
until the required optical thickness is reached.
The reflection and transmission properties of the whole atmosphere are calculated
by combining the properties 
of the individual layers using the 'adding'-equations \citep[see][]{deh87} .

We use a special version of the doubling-adding algorithm, 
namely the so-called 'internal sources' algorithm of \citet{wau94}, which fully includes multiple 
scattering of thermal radiation emitted by the planet's atmosphere.
In this version, the temperature of each atmospheric layer (see Fig.~1) determines the 
amount of thermal radiation that is emitted in the layer. The emission itself is
isotropic. However, scattering of this radiation within the atmospheric layer and/or
within other layers will influence the angular distribution of the radiation as
it emerges at the top of the atmosphere.
We calculate the emerging thermal radiation along twenty angles between 0$^\circ$
(towards the zenith) and 90$^\circ$ (parallel to the atmosphere). Note that since
our atmospheric layers are horizontally homogeneous, the emerging thermal radiation is 
independent of the azimuth angle. Next, assuming the model planet is locally
plane-parallel (thus that the mean-free path of the photons is small enough to ignore atmospheric curvature), we calculate disc-averaged thermal emission spectra by 
integrating the locally emitted spectra over the disc as follows
\begin{equation}
\langle I_\nu \rangle = 2 \int_0^1 \mu I_\nu d\mu
\end{equation}

The cosine of the emission angle ($\mu$) is included in the integration to account
for the spherical shape of the planet.

To calculate the scattering and absorption of radiation with wavelength $\lambda$
within each atmospheric layer, the doubling-adding algorithm needs to know for 
each layer and at the given $\lambda$: its 
extinction optical thickness (the sum of the scattering and absorption 
optical thicknesses), and the single-scattering albedo (the ratio of the scattering
optical thickness to the extinction optical thickness) and the scattering phase
function\footnote{In case polarisation is included, the 4~$\times$~4 phase matrix is
required, see e.g. \citet{deh87}} of the mixture of 
particles and gas molecules in the layer. 

Our model atmospheres not only contain the \textsc{drift-phoenix} particles, but also 
gas molecules. The altitude variation of the different atmospheric gases is produced
by the \textsc{drift-phoenix} code. The absorption coefficients for H$_2$O and CO were
calculated using the HITEMP database \citep{rot10} assuming Voigt line shapes. 
Other molecules that have strong spectral features in the near-infrared, such as CO$_2$ and CH$_4$ are predicted by \textsc{drift-phoenix} to be present in only very small volume mixing ratios 
(generally less than 10$^{-6}$ for CO$_2$ and several orders of magnitude less for CH$_4$),
and should affect the calculated emission spectra only marginally.
Because the focus of this paper is on the scattering effects of clouds, and not so much
on identifying gas signatures or comparisons with observations, we ignored the absorption 
coefficients of these low-concentration gases. 

To efficiently include the gaseous absorption in our calculations of emission spectra,
we used the correlated-$k$ method \citep{lac91}. We first calculated the absorption coefficients of the main gases at a high spectral resolution
at a grid of 20 temperatures between 500-4000 K and 20 pressures between 10$^{-9}$ - 100 bar, which cover the relevant temperature-pressure space for hot exoplanetary atmospheres. 
From these high-spectral resolution spectra, we calculated and tabulated correlated 
$k$-distribution coefficients that we interpolated to the temperature
and pressure of each atmospheric layer.
Also included in our calculations is collision-induced absorption by H$_2$-H$_2$,
using the absorption coefficients of \citet{bor01,bor02}, as well as Rayleigh 
scattering by H$_2$-H$_2$, for which we use a depolarisation factor of 0.02 \citep{irw09}.

\begin{figure}[htp]
\centering
\resizebox{\hsize}{!}{\includegraphics{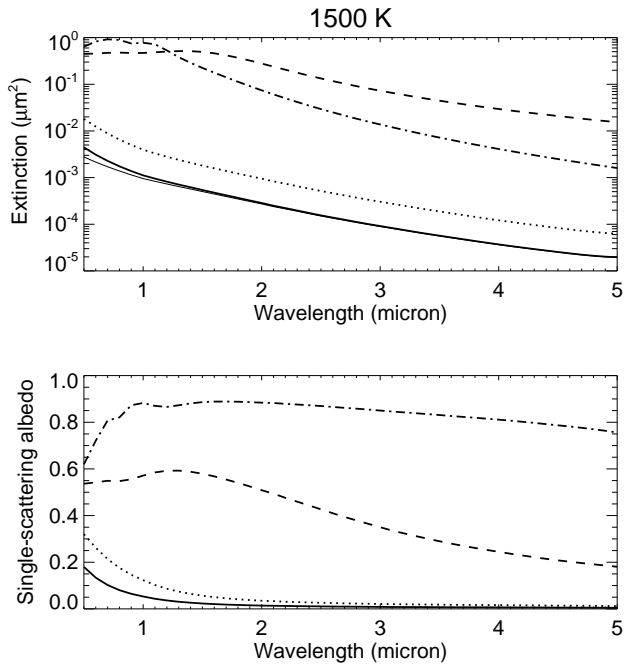}}
\caption{Extinction cross-sections and single-scattering albedos of the four particle types 
         for the 1500~K model atmosphere: the high silicate haze (solid lines); 
         the low silicate haze (dotted lines); the iron cloud (dashed lines);
         and the Al$_2$O$_3$ cloud (dot-dashed lines). 
         The thin solid line in the top panel shows the small-particle approximation 
         for the high silicate haze, which agrees well with the results obtained
         using Mie calculations.}
\label{fig.xsc1}
\end{figure}

\begin{figure}[htp]
\centering
\resizebox{\hsize}{!}{\includegraphics{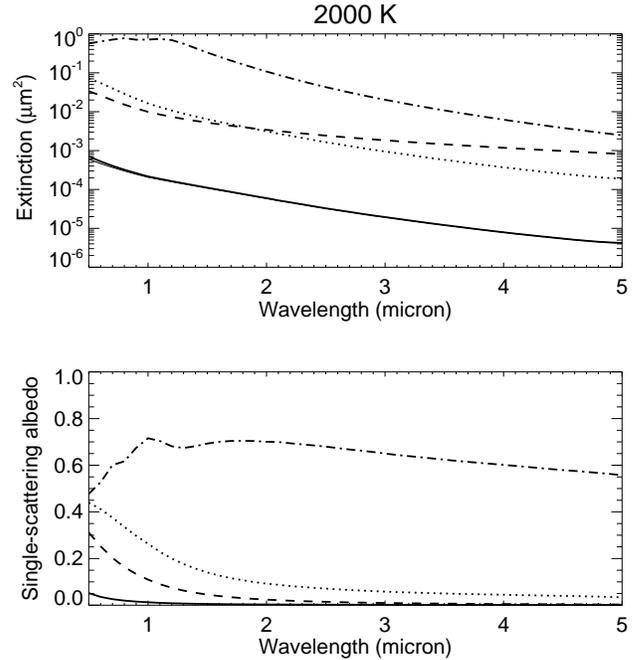}}
\caption{As Fig.~\ref{fig.xsc1}, except for the 2000~K model atmosphere.}
\label{fig.xsc2}
\end{figure}

We calculated the wavelength dependent refractive indices of the various types of 
dirty particles that occur in our model atmospheres from the refractive indices of 
their constituents using 
effective medium theory \citep{bos00}. Then, for each type of particle, we calculated its 
extinction cross-section (i.e.~the surface area with which one particle affects the light), 
its single-scattering albedo and phase function across the required spectral range 
using Mie theory \citep{der84}, assuming spherical particles with radii as given in Table~1.
Calculated extinction cross-sections and single-scattering albedos 
of the particle types occuring in our two model 
atmospheres are shown in Fig.~\ref{fig.xsc1} for the 1500~K atmosphere and in 
Fig.\ref{fig.xsc2} for the 2000~K atmosphere. 
As can be seen in these figures,
none of the constituents of the dirty particles show any noticable absorption features 
below 5 $\mu$m, which results in extinction cross-sections that vary smoothly with the 
wavelength. 
The Al$_2$O$_3$ particles, although only micron-sized, appear to have fairly large 
extinction cross-sections as well as single-scattering albedos at near-infrared 
wavelengths. At these wavelengths, they can thus contribute significantly to 
the scattering of radiation.
The other types of particles appear to scatter less radiation. In particular the
iron particle types have relatively small single-scattering albedos, while the silicate
particle types have small albedos and small extinction cross-sections.
For the silicate haze particles, the extinction cross-section as calculated using 
the small-particle approximation (see Section~2.1) also is plotted and can be seen to 
agree well with the results of the Mie-calculations, which convices us that our approximation
of keeping the extinction cross-sections of these particles fixed and not only varying their 
optical thicknesses is valid.

Comparing Figs.~\ref{fig.xsc1} and~\ref{fig.xsc2}, we see that the particles in the
$T_{\mathrm{eff}}$=2000~K atmosphere have lower albedos and are thus more absorbing
than the particles in the $T_{\mathrm{eff}}$=1500~K atmosphere. The reason is mostly that the former are smaller than the latter (see Fig.~\ref{fig.atm}).

The product of the particle number density, its extinction cross-section at a given
wavelength, and the geometric thickness of the atmospheric layer (assuming a homogeneous
particle number density across the layer), yields the layer's extinction
optical thickness at that wavelength. By multiplying this with the single-scattering albedo
of the particles at the same wavelength, we obtain the layer's scattering optical thickness.
The difference between the layer's extinction optical thickness and
its scattering optical thickness is the layer's absorption optical 
thickness.

%%%%%%%%%%%%%%%%%%%%%%%%%%%%%%%%%%%%%%%%%%%%%%%%%%%%%%%%%%%%%%%%%%%%%%%%%%%%%%%%%%%%%%%%%%%%%%%%%

\section{Results}

\begin{figure}[htp]
\centering
\resizebox{\hsize}{!}{\includegraphics{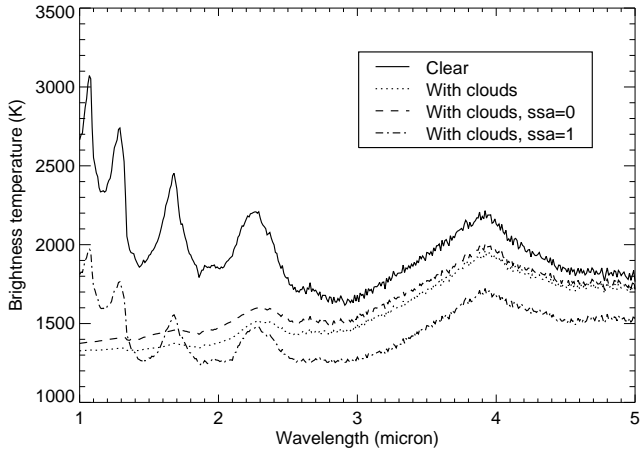}}
\caption{Brightness temperature spectra for the 1500 K case. The solid line indicates the \textsc{drift phoenix} atmosphere without the particles included. For the dotted line the scattering \textsc{drift-phoenix} particles are included. The dashed line shows the spectrum for purely absorbing particle, but same optical thickness (single-scattering albedo, or ssa in the legend, of zero). The dot-dashed line corresponds to the same particles, but now assumed perfectly scattering (single-scattering albedo of unity).}
\label{fig.spec1}
\end{figure}

Thermal emission spectra for the 1500 K case are shown in Fig.~\ref{fig.spec1}. The units plotted here are brightness temperature, since this does not depend on specific planet parameters like radius or distance to the observer. Brightness temperature is the temperature corresponding to a black-body having the radiance that is observed or modeled. Brightness temperature has the additional benefit that it roughly shows where in the atmosphere the radiance is coming from, since it can be compared with the real temperatures in the atmosphere. First a spectrum was calculated for an atmosphere with the \textsc{drift-phoenix} temperature profile and gas abundances, but without the clouds included. Although now the temperature profile is not self-consistent in the \textsc{drift-phoenix} framework, it will show how much including the clouds alone changes the emission spectrum, given a certain temperature profile. The spectrum clearly shows the absorption features of the water vapour, with wavelengths of low opacity giving higher brightness temperatures, because the thermal emission originates lower in the atmosphere. 

Adding in the scattering particles has a very strong effect on the spectrum. In the 1500 K case, the optical thickness of the particles is very large, with a total optical thickness of $\sim$12 at a wavelength of 1 $\mu$m. As usual with small particles, the extinction optical thickness decreases with increasing wavelength (see Fig.~\ref{fig.xsc1}). Hence, the effect of the particles on the spectrum is strongest at short wavelengths. The optical thickness of the particles is larger than that of the water in the windows of low absorption. If the particles are absorbing, this means that most of the radiation will come from the particles themselves at these wavelengths, instead of the water lower in the atmosphere. Higher altitudes correspond here to lower temperatures, so the brightness temperature is reduced,as can be seen in Fig.~\ref{fig.spec1}. Because a more limited altitude range (and thus a more limited temperature range in this case) is probed by the spectrum, the presence of the particles reduces the amplitude of the water features in the spectrum.

We also calculated a spectrum with the single-scattering albedo of all particles set to zero and optical thicknesses are kept identical. This simulates the case when extinction is treated as pure absorption and scattering is neglected. The difference between the dashed and dotted lines thus shows what the effect of the scattering is on the spectrum.  In this case, the water features are slightly more subdued and the entire spectrum shifts by roughly 100 K in brightness temperature at low wavelengths compared to the scattering calculation. The effect of the scattering is two-fold, corresponding to the two terms on the right-hand side of Eq.~\ref{eq.source2}. Firstly, it allows more heat from the lower atmosphere to pass through than expected from extinction only. At the same time, the scattering particles locally reduce the emissivity of the atmosphere, leading to less radiation being emitted by that region of the atmosphere. Scattering can thus increase or reduce the emitted radiance compared to pure absorbing particles, depending on whether the additional scattered radiation is more or less than the reduced emission.

\begin{figure}[htp]
\centering
\resizebox{\hsize}{!}{\includegraphics{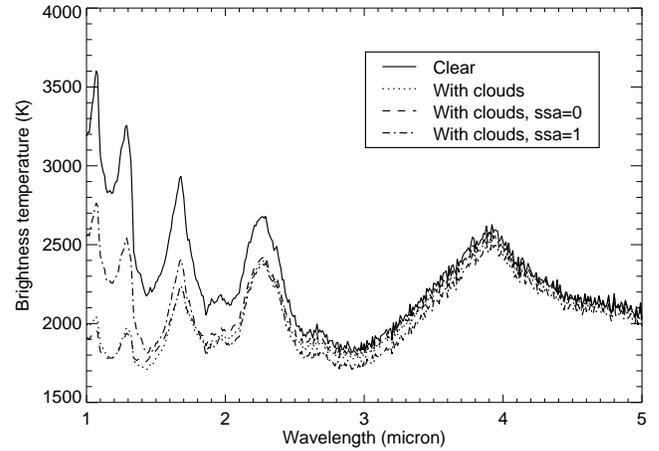}}
\caption{As Fig.~\ref{fig.spec1}, but now for the 2000 K case.}
\label{fig.spec2}
\end{figure}

The spectra for the 2000 K case are shown in Fig.~\ref{fig.spec2}. Optical thicknesses of the cloud particles are lower for this case, which strongly reduces the effect of the particles on the spectrum. Particles also are generally more absorbing in this case, further reducing the effect of scattering. The difference between the spectrum that includes scattering and that without scattering is therefore small. 

In the above spectra the effect of scattering by the particles is significant, but not very large. This is because \textsc{drift-phoenix} predicts small particles with a significant iron content throughout the upper atmosphere, making them relatively absorbing. For the 1500 K case, particle extinction optical depth of unity is reached within the lower silicate haze layer (dotted lines in Fig.~\ref{fig.xsc1}), with significant extinction by the layers above this. Hence, the atmosphere that is emitting the thermal radiation is not very scattering in the \textsc{drift-phoenix} model. However, particles might be larger in other models \citep[see][, their Fig.~5]{hel08}, as well as in real atmospheres. Mie theory shows that larger particles of the same composition will generally have higher single-scattering albedos and more forward-scattering phase functions than small particles. This effect of the particle size also is  visible in the single-scattering albedos in Figs.~\ref{fig.xsc1} and \ref{fig.xsc2}. The more forward-scattering particles with higher single-scattering albedos  would result in more radiation from lower altitudes being scattered upward into space. Furthermore, \textsc{drift-phoenix} has a significant iron content in its particles in the upper atmosphere, which is strongly absorbing. If the iron content is lowered, the single scattering albedo also quickly rises. To illustrate the effect of highly scattering particles, we also calculated a case where the particle single-scattering albedos are set to unity (i.e.~non-absorbing particles). This is the extreme case, but it is realistic for particles larger than the wavelength with a low iron content. For ease of comparison between the cases with different single-scattering albedos, the phase functions and optical thicknesses are kept identical here. Figs.~\ref{fig.spec1} and \ref{fig.spec2} show this case as well (dot-dashed lines). It is clear that the scattering makes a very large difference in how the spectrum looks here: gas absorption features are much more apparent and overal brightness temperature levels are changed substantially. Also note that the relative strengths of the water bands can change significantly by changing the single-scattering albedo of the particles. Thus, if clouds would not be considered in the interpretation of this thermal emission spectrum, one could derive completely different temperature profiles and gas adundances than what is actually present in the atmosphere. If phase functions also would be more forward scattering, as is generally the case with larger particles, this difference with non-scattering particles will be even more pronounced. 

\begin{figure}[htp]
\centering
\resizebox{\hsize}{!}{\includegraphics{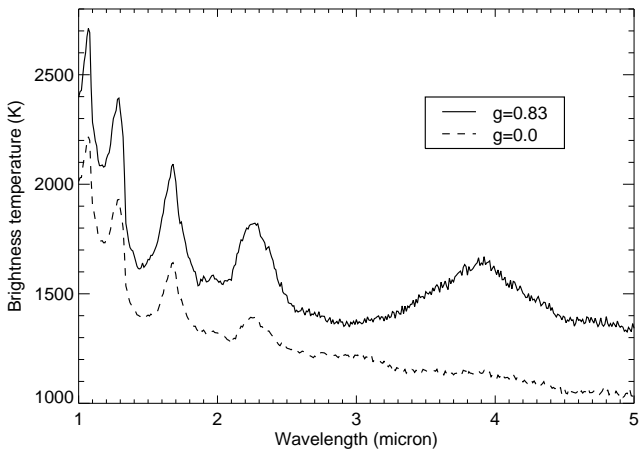}}
\caption{Spectra for two atmospheres with temperatures and optical thicknesses as the 1500 K case at 1 $\mu$m (now taken to be constant with wavelength) and single-scattering albedos of unity, with a difference in phase function. The solid line is the spectrum for the more forward-scattering particles ($g=0.83$), the dashed line is for the isotropically scattering particles ($g = 3 \cdot 10^{-4}$).}
\label{fig.gtest}
\end{figure}

To illustrate the potential effect of a more forward-scattering phase function, we calculate two spectra with only a difference in the phase function. The temperature and optical thicknesses at 1 micron are taken from the 1500 K case, but instead of the particle wavelength-dependence of Fig.~\ref{fig.xsc1} we assume optical thickness to be constant with wavelength and single-scattering albedos of unity, as is common for large non-absorbing particles. Also, all particles throughout the atmosphere were assumed identical. The two different phase functions were also assumed constant with wavelength and were taken from the T$_{\mathrm{eff}} =$  1500 K Al$_2$O$_3$ cloud at a wavelength of 0.3 micron (asymmetry parameter of $g=0.83$) and the T$_{\mathrm{eff}} =$2000 K high silicate haze at 5 micron ($g = 3 \cdot 10^{-4}$). The first is thus strongly forward-scattering, whereas the latter is almost isotropically scattering. The two spectra with the two different particles are shown in Fig.~\ref{fig.gtest}. Even though the particle optical thicknesses for both cases are identical, the spectra differ very strongly. The forward-scattering particles allow more heat from the warmer regions below the clouds to escape to space, giving higher radiances and stronger absorption features. Hence, assuming isotropic scattering can give errors of many tens of percent on the spectrum if the particles are in fact strongly forward-scattering.

\section{Conclusions and discussion}

Cloud and/or haze particles in the atmospheres of hot exoplanets or brown dwarfs can have a strong effect on their thermal emission spectra, changing their brightness and colour at different wavelengths. These thermal emission spectra can be measured either directly, or from their secondary eclipse. The effect of the clouds is strongest at shorter wavelengths and colder temperatures for the \textsc{drift-phoenix} model, and generally for small particles. This makes the brown dwarf more `red' in the infrared. Although the latter effect is clear from previous studies, not much attention has been given in the past to the contribution from scattered light to the spectrum. Here, we show that not only the extinction of the cloud, but also the scattering nature of the particles can be important in determining the emission spectra for these objects. Scattering can affect the emission spectra especially when the particles are large and have little iron in them. Hence, knowledge of the scattering properties such as single-scattering albedo and phase function can be crucial in calculating accurate spectra of sub-stellar atmospheres. Errors in these parameters will result in errors in atmospheric properties derived from infrared direct measurements or secondary eclipses. Also the self-consistent calculation of the temperature structure in such an atmosphere can thus depend significantly on the assumed or calculated scattering properties in the atmosphere. For instance, Fig.~\ref{fig.gtest} shows that more forward-scattering phase functions allow more heat from lower altitudes to escape to space, leading to an increase in cooling of these lower regions compared to more isotropically scattering particles. This also will affect calculations made using the isotropically scattering \textsc{phoenix} model, if the particles are more scattering than the silicate haze in the two \textsc{drift-phoenix} cases presented here. The errors due to the parameterisation of the phase function in terms of $g$ are bound to be less severe, but probably can reach several percent in the calculated radiance. Thus, care must be taken in the choice of radiative transfer method for heat balance or spectral calculations, depending on the type of particles that are predicted.  

Current analyses of transiting exoplanet emission spectra, as derived from secondary eclipses, suffer from a large degeneracy, with a wide range of atmospheres fitting the limited measurements equally well \citep[e.g.][]{mad09}. This means that atmospheric properties like temperature and composition are poorly constrained by the observations. Most current secondary eclipse measurements are taken in the near-infrared by the warm Spitzer mission and ground-based telescopes and none of the current fits to the measurements include clouds. We show here that adding scattering clouds only will add to the degeneracy, as e.g.~the depths of absorption features or brightness temperatures at a certain wavelength in the near-infrared strongly depend on the scattering properties and optical thicknesses of the particles, as well as the temperature structure and gas abundances. As there is no evidence that the currently analysed transiting exoplanets are cloud-free, clouds should be taken into account when fitting the observations, hence increasing the uncertainty in the other free parameters such as temperature and gas abundances. A cloud physics model like \textsc{drift-phoenix} will be valuable in exploring the parameter space and physically explaining the appearance of the measured spectra. Also, the best-fit temperature profiles and gas adundances, derived without clouds, can be substantially in error when clouds are present in the atmosphere. 

Our spectra also show that the composition of clouds themselves cannot be constrained well from near-infrared measurements of the thermal emission alone, since multiple combinations of composition mixtures and particle sizes can give very similar extinction and scattering properties. Measurements at longer wavelengths, where the expected silicate particles have absorption features, can possibly help to reduce the degeneracy, for both directly detected exoplanets and transiting exoplanets \citep{hel08b}. Also observations of reflected starlight at short wavelengths and transit transmission measurements \citep{lec08,pon08} can help to determine the haze and cloud structure and so more accurately determine the temperature structure and gas adundances from secondary eclipse observations. Note that scattering of the starlight may play a role in transit transmission measurements as well.

The fact that the thermal radiation can be scattered also means that it can be polarised. For a symmetrical, spherical planet the disc-integrated polarisation will be zero, but when the planet or brown dwarf is flattened \citep{sen09,sen10} or horizontally inhomogeneous it can result in the infrared thermal radiation being polarised. Our radiative transfer code also includes an accurate description of the polarisation of the light and so it can calculate polarisation of the light given a certain viewing geometry. In a separate work we will investigate the possible polarisation signals from the thermal radiation of horizontally inhomogeneous planets.

Finally, it also is interesting to note that the small dark particles predicted by \textsc{drift-phoenix} also will result in a lower geometric albedo of the planet at visible wavelengths compared to Rayleigh scattering \citep{bur08a} or relatively large enstatite particles \citep{sud00,hoo08}. The latter two do not model the cloud physics, but assume an ad hoc particle size distribution. The radiative transfer in \citet{hoo08} differs from other work cited here, as it takes into account all three spatial dimensions of the atmosphere. Because the small iron-rich particles are dark, this haze may contribute to the low albedos of some hot Jupiters \citep[e.g.][]{row08}. The darkening effect of small iron particles has already been noted by \citet{sea00}, but the \textsc{drift-phoenix} model might provide a physical basis for this argument.

\begin{acknowledgements}
We would like to thank Peter Hauschildt, Mark Marley and Adam Burrows for answering questions regarding their radiative transfer methods. We thank Patrick Irwin for interesting discussion regarding scattering of thermal radiation. We thank the referee, France Allard, for useful comments. This research is supported by the Netherlands Organisation for Scientific Research (NWO).
\end{acknowledgements}


\begin{thebibliography}{57}
\expandafter\ifx\csname natexlab\endcsname\relax\def\natexlab#1{#1}\fi

\bibitem[{{Ackerman} \& {Marley}(2001)}]{ack01}
{Ackerman}, A.~S. \& {Marley}, M.~S. 2001, ApJ, 556, 872

\bibitem[{{Allard} \& {Hauschildt}(1995)}]{all95}
{Allard}, F. \& {Hauschildt}, P.~H. 1995, ApJ, 445, 433

\bibitem[{{Allard} {et~al.}(2001){Allard}, {Hauschildt}, {Alexander},
  {Tamanai}, \& {Schweitzer}}]{all01}
{Allard}, F., {Hauschildt}, P.~H., {Alexander}, D.~R., {Tamanai}, A., \&
  {Schweitzer}, A. 2001, ApJ, 556, 357

\bibitem[{{Barman} {et~al.}(2001){Barman}, {Hauschildt}, \& {Allard}}]{bar01}
{Barman}, T.~S., {Hauschildt}, P.~H., \& {Allard}, F. 2001, ApJ, 556, 885

\bibitem[{{Barman} {et~al.}(2005){Barman}, {Hauschildt}, \& {Allard}}]{bar05}
{Barman}, T.~S., {Hauschildt}, P.~H., \& {Allard}, F. 2005, ApJ, 632, 1132

\bibitem[{{Bonnefoy} {et~al.}(2010){Bonnefoy}, {Chauvin}, {Rojo}, {Allard},
  {Lagrange}, {Homeier}, {Dumas}, \& {Beuzit}}]{bon10}
{Bonnefoy}, M., {Chauvin}, G., {Rojo}, P., {et~al.} 2010, A\&A, 512, A52+

\bibitem[{{Borysow}(2002)}]{bor02}
{Borysow}, A. 2002, A\&A, 390, 779

\bibitem[{{Borysow} {et~al.}(2001){Borysow}, {Jorgensen}, \& {Fu}}]{bor01}
{Borysow}, A., {Jorgensen}, U.~G., \& {Fu}, Y. 2001, J. Quant. Spectro. Rad.
  Trans., 68, 235

\bibitem[{{Bosch} {et~al.}(2000){Bosch}, {Ferre-Borrull}, {Leinfellner}, \&
  {Canillas}}]{bos00}
{Bosch}, S., {Ferre-Borrull}, J., {Leinfellner}, N., \& {Canillas}, A. 2000,
  Surface Science, 453, 9

\bibitem[{{Bowler} {et~al.}(2010){Bowler}, {Liu}, {Dupuy}, \&
  {Cushing}}]{bow10}
{Bowler}, B.~P., {Liu}, M.~C., {Dupuy}, T.~J., \& {Cushing}, M.~C. 2010, ApJ,
  723, 850

\bibitem[{{Burgasser}(2009)}]{burg09}
{Burgasser}, A.~J. 2009, {Mem.~S.A.It.}, 80, 658

\bibitem[{{Burrows} {et~al.}(2008{\natexlab{a}}){Burrows}, {Budaj}, \&
  {Hubeny}}]{bur08}
{Burrows}, A., {Budaj}, J., \& {Hubeny}, I. 2008{\natexlab{a}}, ApJ, 678, 1436

\bibitem[{{Burrows} {et~al.}(2008{\natexlab{b}}){Burrows}, {Ibgui}, \&
  {Hubeny}}]{bur08a}
{Burrows}, A., {Ibgui}, L., \& {Hubeny}, I. 2008{\natexlab{b}}, ApJ, 682, 1277

\bibitem[{{Burrows} {et~al.}(2006){Burrows}, {Sudarsky}, \& {Hubeny}}]{bur06}
{Burrows}, A., {Sudarsky}, D., \& {Hubeny}, I. 2006, ApJ, 640, 1063

\bibitem[{{Carlson} {et~al.}(1993){Carlson}, {Lacis}, \& {Rossow}}]{car93}
{Carlson}, B.~E., {Lacis}, A.~A., \& {Rossow}, W.~B. 1993, J. Geophys. Res.,
  98, 5251

\bibitem[{{Chandrasekhar}(1960)}]{cha60}
{Chandrasekhar}, S. 1960, {Radiative transfer}, ed. {Chandrasekhar, S.}

\bibitem[{{Croll} {et~al.}(2010){Croll}, {Jayawardhana}, {Fortney},
  {Lafreni{\`e}re}, \& {Albert}}]{cro10}
{Croll}, B., {Jayawardhana}, R., {Fortney}, J.~J., {Lafreni{\`e}re}, D., \&
  {Albert}, L. 2010, ApJ, 718, 920

\bibitem[{{Cushing} {et~al.}(2006){Cushing}, {Roellig}, {Marley}, {Saumon},
  {Leggett}, {Kirkpatrick}, {Wilson}, {Sloan}, {Mainzer}, {Van Cleve}, \&
  {Houck}}]{cus06}
{Cushing}, M.~C., {Roellig}, T.~L., {Marley}, M.~S., {et~al.} 2006, ApJ, 648,
  614

\bibitem[{{de Haan} {et~al.}(1987){de Haan}, {Bosma}, \& {Hovenier}}]{deh87}
{de Haan}, J.~F., {Bosma}, P.~B., \& {Hovenier}, J.~W. 1987, A\&A, 183, 371

\bibitem[{{de Rooij} \& {van der Stap}(1984)}]{der84}
{de Rooij}, W.~A. \& {van der Stap}, C.~C.~A.~H. 1984, A\&A, 131, 237

\bibitem[{{Deming} \& {Seager}(2009)}]{dem09}
{Deming}, D. \& {Seager}, S. 2009, Nat, 462, 301

\bibitem[{{Grinspoon} {et~al.}(1993){Grinspoon}, {Pollack}, {Sitton},
  {Carlson}, {Kamp}, {Baines}, {Encrenaz}, \& {Taylor}}]{gri93}
{Grinspoon}, D.~H., {Pollack}, J.~B., {Sitton}, B.~R., {et~al.} 1993, Plan. \&
  Space Sci., 41, 515

\bibitem[{{Hanel} {et~al.}(2003){Hanel}, {Conrath}, {Jennings}, \&
  {Samuelson}}]{han03}
{Hanel}, R.~A., {Conrath}, B.~J., {Jennings}, D.~E., \& {Samuelson}, R.~E.
  2003, {Exploration of the Solar System by Infrared Remote Sensing: Second
  Edition} (pp.~534.~ISBN 0521818974.~Cambridge, UK: Cambridge University
  Press)

\bibitem[{{Hansen} \& {Travis}(1974)}]{han74}
{Hansen}, J.~E. \& {Travis}, L.~D. 1974, Space Sci. Rev., 16, 527

\bibitem[{{Hauschildt}(1992)}]{hau92}
{Hauschildt}, P.~H. 1992, J. Quant. Spectro. Rad. Trans., 47, 433

\bibitem[{{Hauschildt} \& {Baron}(2006)}]{hau06}
{Hauschildt}, P.~H. \& {Baron}, E. 2006, A\&A, 451, 273

\bibitem[{{Helling} {et~al.}(2008{\natexlab{a}}){Helling}, {Ackerman},
  {Allard}, {Dehn}, {Hauschildt}, {Homeier}, {Lodders}, {Marley}, {Rietmeijer},
  {Tsuji}, \& {Woitke}}]{hel08}
{Helling}, C., {Ackerman}, A., {Allard}, F., {et~al.} 2008{\natexlab{a}}, Mon.
  Not. R. Astron. Soc., 391, 1854

\bibitem[{{Helling} {et~al.}(2008{\natexlab{b}}){Helling}, {Dehn}, {Woitke}, \&
  {Hauschildt}}]{hel08a}
{Helling}, C., {Dehn}, M., {Woitke}, P., \& {Hauschildt}, P.~H.
  2008{\natexlab{b}}, ApJ, 675, L105

\bibitem[{{Helling} {et~al.}(2008{\natexlab{c}}){Helling}, {Woitke}, \&
  {Thi}}]{hel08b}
{Helling}, C., {Woitke}, P., \& {Thi}, W. 2008{\natexlab{c}}, A\&A, 485, 547

\bibitem[{{Hood} {et~al.}(2008){Hood}, {Wood}, {Seager}, \& {Collier
  Cameron}}]{hoo08}
{Hood}, B., {Wood}, K., {Seager}, S., \& {Collier Cameron}, A. 2008, Mon. Not.
  R. Astron. Soc., 389, 257

\bibitem[{{Hubeny}(2003)}]{hub03}
{Hubeny}, I. 2003, in Astronomical Society of the Pacific Conference Series,
  Vol. 288, Stellar Atmosphere Modeling, ed. {I.~Hubeny, D.~Mihalas, \&
  K.~Werner}, 17

\bibitem[{{Hubeny} \& {Lanz}(1995)}]{hub95}
{Hubeny}, I. \& {Lanz}, T. 1995, ApJ, 439, 875

\bibitem[{{Irwin}(2009)}]{irw09}
{Irwin}, P.~G.~J. 2009, {Giant Planets of Our Solar System: Atmospheres,
  Composition, and Structure, Springer Praxis Books, Volume .~ISBN
  978-3-540-85157-8.~Springer Berlin Heidelberg, 2009}

\bibitem[{{Janson} {et~al.}(2010){Janson}, {Bergfors}, {Goto}, {Brandner}, \&
  {Lafreni{\`e}re}}]{jan10}
{Janson}, M., {Bergfors}, C., {Goto}, M., {Brandner}, W., \& {Lafreni{\`e}re},
  D. 2010, ApJ, 710, L35

\bibitem[{{Knutson} {et~al.}(2009){Knutson}, {Charbonneau}, {Cowan}, {Fortney},
  {Showman}, {Agol}, {Henry}, {Everett}, \& {Allen}}]{knu09}
{Knutson}, H.~A., {Charbonneau}, D., {Cowan}, N.~B., {et~al.} 2009, ApJ, 690,
  822

\bibitem[{Lacis \& Oinas(1991)}]{lac91}
Lacis, A.~A. \& Oinas, V. 1991, J. Geophys. Res., 96, 9027

\bibitem[{{Lecavelier Des Etangs} {et~al.}(2008){Lecavelier Des Etangs},
  {Pont}, {Vidal-Madjar}, \& {Sing}}]{lec08}
{Lecavelier Des Etangs}, A., {Pont}, F., {Vidal-Madjar}, A., \& {Sing}, D.
  2008, A\&A, 481, L83

\bibitem[{{Leggett} {et~al.}(1998){Leggett}, {Allard}, \& {Hauschildt}}]{leg98}
{Leggett}, S.~K., {Allard}, F., \& {Hauschildt}, P.~H. 1998, ApJ, 509, 836

\bibitem[{{Madhusudhan} \& {Seager}(2009)}]{mad09}
{Madhusudhan}, N. \& {Seager}, S. 2009, ApJ, 707, 24

\bibitem[{{Patience} {et~al.}(2010){Patience}, {King}, {de Rosa}, \&
  {Marois}}]{pat10}
{Patience}, J., {King}, R.~R., {de Rosa}, R.~J., \& {Marois}, C. 2010, A\&A,
  517, A76+

\bibitem[{{Pont} {et~al.}(2008){Pont}, {Knutson}, {Gilliland}, {Moutou}, \&
  {Charbonneau}}]{pon08}
{Pont}, F., {Knutson}, H., {Gilliland}, R.~L., {Moutou}, C., \& {Charbonneau},
  D. 2008, Mon. Not. R. Astron. Soc., 385, 109

\bibitem[{{Rothman} {et~al.}(2010){Rothman}, {Gordon}, {Barber}, {Dothe},
  {Gamache}, {Goldman}, {Perevalov}, {Tashkun}, \& {Tennyson}}]{rot10}
{Rothman}, L.~S., {Gordon}, I.~E., {Barber}, R.~J., {et~al.} 2010, J. Quant.
  Spectro. Rad. Trans., 111, 2139

\bibitem[{{Rowe} {et~al.}(2008){Rowe}, {Matthews}, {Seager}, {Miller-Ricci},
  {Sasselov}, {Kuschnig}, {Guenther}, {Moffat}, {Rucinski}, {Walker}, \&
  {Weiss}}]{row08}
{Rowe}, J.~F., {Matthews}, J.~M., {Seager}, S., {et~al.} 2008, ApJ, 689, 1345

\bibitem[{{Seager} {et~al.}(2000){Seager}, {Whitney}, \& {Sasselov}}]{sea00}
{Seager}, S., {Whitney}, B.~A., \& {Sasselov}, D.~D. 2000, ApJ, 540, 504

\bibitem[{{Sengupta} \& {Marley}(2009)}]{sen09}
{Sengupta}, S. \& {Marley}, M.~S. 2009, ApJ, 707, 716

\bibitem[{{Sengupta} \& {Marley}(2010)}]{sen10}
{Sengupta}, S. \& {Marley}, M.~S. 2010, ApJ, 722, L142

\bibitem[{{Sing} {et~al.}(2009){Sing}, {D{\'e}sert}, {Lecavelier Des Etangs},
  {Ballester}, {Vidal-Madjar}, {Parmentier}, {Hebrard}, \& {Henry}}]{sin09}
{Sing}, D.~K., {D{\'e}sert}, J., {Lecavelier Des Etangs}, A., {et~al.} 2009,
  A\&A, 505, 891

\bibitem[{{Sudarsky} {et~al.}(2000){Sudarsky}, {Burrows}, \& {Pinto}}]{sud00}
{Sudarsky}, D., {Burrows}, A., \& {Pinto}, P. 2000, ApJ, 538, 885

\bibitem[{{Swain} {et~al.}(2009){Swain}, {Tinetti}, {Vasisht}, {Deroo},
  {Griffith}, {Bouwman}, {Chen}, {Yung}, {Burrows}, {Brown}, {Matthews},
  {Rowe}, {Kuschnig}, \& {Angerhausen}}]{swai09}
{Swain}, M.~R., {Tinetti}, G., {Vasisht}, G., {et~al.} 2009, ApJ, 704, 1616

\bibitem[{{Toon} {et~al.}(1989){Toon}, {McKay}, {Ackerman}, \&
  {Santhanam}}]{too89}
{Toon}, O.~B., {McKay}, C.~P., {Ackerman}, T.~P., \& {Santhanam}, K. 1989, J.
  Geophys. Res., 94, 16287

\bibitem[{{Tsang} {et~al.}(2008){Tsang}, {Irwin}, {Wilson}, {Taylor}, {Lee},
  {de Kok}, {Drossart}, {Piccioni}, {Bezard}, \& {Calcutt}}]{tsa08}
{Tsang}, C.~C.~C., {Irwin}, P.~G.~J., {Wilson}, C.~F., {et~al.} 2008, Journal
  of Geophysical Research (Planets), 113, E12, doi:10.1029/2008JE003089

\bibitem[{{Tsuji} {et~al.}(1996){Tsuji}, {Ohnaka}, {Aoki}, \&
  {Nakajima}}]{tsu96}
{Tsuji}, T., {Ohnaka}, K., {Aoki}, W., \& {Nakajima}, T. 1996, A\&A, 308, L29

\bibitem[{{Wauben} {et~al.}(1994){Wauben}, {de Haan}, \& {Hovenier}}]{wau94}
{Wauben}, W.~M.~F., {de Haan}, J.~F., \& {Hovenier}, J.~W. 1994, A\&A, 282, 277

\bibitem[{{Witte} {et~al.}(2009){Witte}, {Helling}, \& {Hauschildt}}]{wit09}
{Witte}, S., {Helling}, C., \& {Hauschildt}, P.~H. 2009, A\&A, 506, 1367

\bibitem[{{Woitke} \& {Helling}(2003)}]{woi03}
{Woitke}, P. \& {Helling}, C. 2003, A\&A, 399, 297

\bibitem[{{Woitke} \& {Helling}(2004)}]{woi04}
{Woitke}, P. \& {Helling}, C. 2004, A\&A, 414, 335

\bibitem[{{Wood} {et~al.}(2011){Wood}, {Maxted}, {Smalley}, \& {Iro}}]{woo11}
{Wood}, P.~L., {Maxted}, P.~F.~L., {Smalley}, B., \& {Iro}, N. 2011, Mon. Not.
  R. Astron. Soc., 171

\end{thebibliography}
\end{document}